\documentclass[a4paper,12pt]{article}
\usepackage[utf8]{inputenc}
\usepackage{amsfonts}
\usepackage{amssymb}
\usepackage{graphicx}
\usepackage{amsmath}
\usepackage[table,xcdraw]{xcolor}
\usepackage{tikz}
\usetikzlibrary{decorations.markings}
\usepackage{enumerate}
\usepackage{ulem}
\usepackage{cases}
\usepackage{subfig}
\usepackage{pgfplots}
\usepackage{color}
\usepackage{tikz}\usetikzlibrary{calc}
\usepackage{float}
\usepackage{here}
\usepackage{cite}
\usepackage{mathrsfs}
\usepackage{float,epsfig}
\usepackage{dcolumn}

\usepackage{graphicx}
\usepackage{bm}
\usepackage{amsmath,amssymb,amsthm}
\usepackage[colorlinks=true,linkcolor=blue,citecolor=red]{hyperref}
\newcommand{\be}{\begin{equation}}
\newcommand{\ee}{\end{equation}}
\newcommand{\bea}{\setlength\arraycolsep{2pt} \begin{eqnarray}}
\newcommand{\eea}{\end{eqnarray}}

\def\0{{\sst{(0)}}}
\def\1{{\sst{(1)}}}
\def\2{{\sst{(2)}}}
\def\3{{\sst{(3)}}}
\def\4{{\sst{(4)}}}
\def\5{{\sst{(5)}}}
\def\6{{\sst{(6)}}}
\def\7{{\sst{(7)}}}
\def\8{{\sst{(8)}}}
\def\sst#1{{\scriptscriptstyle #1}}

\newcommand{\pgftextcircled}[1]{
    \setbox0=\hbox{#1}%
    \dimen0\wd0%
    \divide\dimen0 by 2%
    \begin{tikzpicture}[baseline=(a.base)]%
        \useasboundingbox (-\the\dimen0,0pt) rectangle (\the\dimen0,1pt);
        \node[circle,draw,outer sep=0pt,inner sep=0.1ex] (a) {#1};
    \end{tikzpicture}
}

\definecolor{lime}{HTML}{A6CE39}

\makeatletter \@addtoreset{equation}{section}

\usepackage{multirow}
\newcommand*\frontaleye{%
       \scalebox{0.25}{
\tikzset{every picture/.style={line width=0.75pt}}
\begin{tikzpicture}[x=0.75pt,y=0.75pt,yscale=-1,xscale=1]
\draw  [draw opacity=0][fill={rgb, 255:red, 140; green, 196; blue, 74 }  ,fill opacity=1 ] (104.5,179.99) .. controls (104.5,171.39) and (111.48,164.41) .. (120.08,164.41) .. controls (128.68,164.41) and (135.66,171.39) .. (135.66,179.99) .. controls (135.66,188.6) and (128.68,195.57) .. (120.08,195.57) .. controls (111.48,195.57) and (104.5,188.6) .. (104.5,179.99) -- cycle ;
\draw  [fill={rgb, 255:red, 0; green, 0; blue, 0 }  ,fill opacity=1 ] (84.83,179.7) .. controls (84.92,169.21) and (100.66,160.85) .. (119.99,161.01) .. controls (139.32,161.18) and (154.92,169.81) .. (154.83,180.3) .. controls (152.49,171.04) and (137.8,163.81) .. (119.97,163.66) .. controls (102.13,163.51) and (87.32,170.48) .. (84.83,179.7) -- cycle ;
\draw  [fill={rgb, 255:red, 0; green, 0; blue, 0 }  ,fill opacity=1 ] (113.6,179.91) .. controls (113.6,176.19) and (116.63,173.16) .. (120.36,173.16) .. controls (124.08,173.16) and (127.11,176.19) .. (127.11,179.91) .. controls (127.11,183.64) and (124.08,186.67) .. (120.36,186.67) .. controls (116.63,186.67) and (113.6,183.64) .. (113.6,179.91) -- cycle ;
\draw  [fill={rgb, 255:red, 0; green, 0; blue, 0 }  ,fill opacity=1 ] (154.65,179.7) .. controls (154.65,190.58) and (139.02,199.41) .. (119.74,199.41) .. controls (100.46,199.41) and (84.83,190.58) .. (84.83,179.7) .. controls (87.23,189.3) and (101.95,196.67) .. (119.74,196.67) .. controls (137.53,196.67) and (152.25,189.3) .. (154.65,179.7) -- cycle ;
\draw  [draw opacity=0][fill={rgb, 255:red, 255; green, 255; blue, 255 }  ,fill opacity=1 ] (125.45,172.24) .. controls (125.45,170.4) and (126.94,168.91) .. (128.78,168.91) .. controls (130.62,168.91) and (132.11,170.4) .. (132.11,172.24) .. controls (132.11,174.08) and (130.62,175.57) .. (128.78,175.57) .. controls (126.94,175.57) and (125.45,174.08) .. (125.45,172.24) -- cycle ;
\end{tikzpicture}
}\kern-.5em}

\begin{document}
\title{{ \textbf{\Large {  Optical Aspect  of    Cosmological  Black Holes in   Einstein-Maxwell-Dilaton  Theory  }}}}
\author{ {\small Hajar Belmahi$^1$\footnote{Corresponding author: hajar\_belmahi@um5.ac.ma},  Amin Mohamed  Rbah$^2$\thanks{
\text{Authors in alphabetical order.}} \hspace*{-8pt}} \\
{\small $^1$Facult\'{e} des Sciences, Universit\'{e} Mohammed V de
Rabat, Rabat, Morocco}\\
{\small   $^2$Faculté Polydisciplinaire de  B\'{e}ni Mellal, Universit\'{e} Sultan Moulay Slimane, B\'{e}ni Mellal, Morocco} }
\maketitle

	\begin{abstract}
Motivated by string theory scenarios, we study the  optical  aspect of AdS  black holes in      Einstein-Maxwell-dilaton  theory.  Concretely, 
  we investigate and examine    the shadows and the deflection angle of light rays by  such  cosmological black holes.     Concerning    the shadows,  we first deal with   the non-rotating   solutions.  As expected, we obtain   perfect circular shadows  where their  sizes are controlled by the involved parameter including the charge and the cosmological constant.  Combining    the  Newman-Janis formalism and  the Hamilton-Jacobi algorithm, we approach the rotating  black hole  shadows  using  one dimensional  real curves.  Among  others,    we  observe that the   size and the  shape shadows  depend  on certain parameters including   the rotating   one.  To  make contact with Event Horizon Telescope  observational  data,   we show that certain constrains should be  imposed on such parameters.  Then,  we  study the behaviors of  the light   rays  near     such   cosmological   black holes by  computing the  deflection angle in terms  of   Einstein-Maxwell-dilaton  theory parameters. Specifically, we reveal that the effect of the cosmological constant on the deflection angle  depends on  the coupling between the black hole parameters.  Introducing the rotating parameter, we observe this effect becomes similar to that of the cosmological constant in ordinary AdS black holes.\\

		{\noindent}
	\end{abstract}
\newpage

\tableofcontents
\newpage

\section{Introduction}
 Black holes in non-trivial physical theories  are  considered as   challenging and interesting subjects that  have received  a remarkable interest.    It has   been a central topic in theoretical physics, offering profound insights into the nature of gravity, quantum mechanics, and cosmology \cite{A0,A1,A2,A3,A4,A5}.  Actually, the subject of black holes is of great importance, 
not only in astrophysics, but also in high  energy  physics theories  including string theory and related topics  explored to  unify general relativity with quantum physics \cite{A7,A61s,primo,JET,C1,A157,A136,A137,A8,A9,A10}. In general, the investigation of black holes is developed from the consideration of two key topics: thermodynamics and optics.

The unexpected link between  the black holes and  the thermodynamics has opened up an intriguing new  field of research in black hole physics. Despite being previously thought to be exclusively Einstein equation solutions, the black holes have been shown to have interesting thermodynamic features that challenge the  understanding of the fundamental nature of the universe. This paradigm change is made possible by the pivotal Hawking area theorem. Hawking was able  to confirm the  Bekenstein  conjecture by establishing a thermodynamic relationship between energy and area \cite{A11,A12,A13,A14}. This  brings a thermodynamic aspect to these enigmatic astronomical objects.

Encouraged by  empirical findings,   including  the detection of the gravitational waves,  it has been 
   provided   a strong evidence of the black hole existence \cite{LIGs,BGRA}.   Such  interesting  observations are supported by the  image of  the black hole  developed by 
Event Horizon  Telescope (EHT) collaboration  groups.  In 2019, this  collaboration made a groundbreaking achievement by releasing the first-ever direct image of a black hole located at the center of the galaxy M87, confirming the picture  predicted by the general theory of relativity \cite{A56,B10,A57}. Subsequently, the EHT  international collaboration  presented another image associated with a supermassive black  hole Sgr A$*$ \cite{A58,A59,A60,A61,A62}. In 2022, the EHT team provided images of the shadow of this black hole, offering valuable information about the size, the mass, and the rotation parameter.  This image could provide data on the black hole shape.

 Motivated by such   observational data,  the optical aspect of various    black holes     has  been largely  studied  using  different analytical and numerical methods.  A close inspection, in such activities, reveals that   two relevant   optical notions  have been  investigated:   the shadow and the  deflection angle of  the light rays  near   a  black hole.  Indeed,  the shadow concept has been approached by  using   the  Hamilton-Jacobi  formalism in different gravity models \cite{Vagnozzi:2022moj,A64,Jafarzade, Yanxx,A154A,A65,A66,A68,A69,A73,A74,A77}.     In this way, the equations of motion of   massless particles  near   several  black holes have been found and  largely examined using algebraic  real geometries.  In four dimensions,  for instance, the shadows have been illustrated   via   one-dimensional  real curves with different geometrical configurations \cite{A78,A79,Lemos,shw,A80,A81}. It has been shown that    the    size and the shape   of  such   curves depend on the black hole  parameter  space.    Concretely,     the shadows of   the non-rotating black holes  have been  found to be  perfect circles  where the size could be controlled by certain parameters such as   the charge.   The rotating   parameter,  however,  deforms  such geometries   producing   non-trivial  configurations known as  D or cardioid shapes \cite{A75,A76,A83, A94,A159,A102,A89}. Effectively, the cardioid shapes have been observed  in the case of   rotating black holes in type IIB superstring and M-theory. In particular,  it has been shown that the shadows of 5D black holes embedded in type IIB superstring geometries exhibit a D-appearance for small values of the D3-brane number.  Moreover,  a possible  transition to cardioid    has been also observed   for large brane number values\cite{A95,A96,A90}. For superentropic black hole shadows in arbitrary dimensions,  this shadow geometry transition  has been studied and illustrated  in a beautiful  manner. It has been proved that  the shadows undergo certain geometric transitions depending on the space-time dimension \cite{sm1,sm2,sm3}.

 In parallel  studies, the deflection angle of light rays  near to black holes has been  also investigated. This angle  could   provide interesting  features   which  could be exploited to impose physical constraints   on  the underling  black hole  gravity models.   Many works on such an  optical  quantity   have been  elaborated using different    computation methods \cite{A117,A118,A119,A120,A121,A122,A123,A124}. Concretely, Gibbons and Werner have proposed a direct method to  calculate the  deflection angle  by means of  the Gauss-Bonnet theorem \cite{A126,A127,BelhajH,A128,A129,Virbhadra,A131,A133}.  This approach relies  on  an  optical  space  metric where the angle is perceived as a partially topological effect that can be calculated by utilizing the Gaussian curvature of the optical metric and employing   weak approximations.
 
 Among the various modifications and extensions of General Relativity, the Einstein-Maxwell-dilaton (EMD) theory is of particular interest due to its significance in string theory and other high-energy frameworks. This theory incorporates a scalar field called the dilaton, which interacts with the electromagnetic field \cite{Hirschmann}. The presence of the dilaton has profound implications for the causal structure and thermodynamic properties of black holes. Furthermore, the study of asymptotically (anti)-de Sitter dilaton black holes has revealed a rich array of intriguing physical phenomena, including  thermodynamics being a particularly well-explored aspect  \cite{Hirschmann, Yu, Al,Lu, Charmousis}. 
 
 The purpose of this paper is to contribute to the study of AdS black holes in EMD theory by investigating the corresponding  optical properties.
   Concretely, 
  we  examine    the shadows and the deflection angle of light rays by  such  cosmological black holes.     Concerning    the shadows,  we first  study the non-rotating   solutions. In particular, we  obtain     perfect circular shadows  where their  sizes are controlled by the involved parameter including the charge and the cosmological constant.  Using   the  Newman-Janis formalism and  the Hamilton-Jacobi algorithm, we  investigate  the rotating  black hole solutions using  real curves.  Among  others,    we  find that the   size and  the shape shadows  depend  on  the rotating  parameter and the remaining  ones.  To  make contact with EHT  observational  data,  we show that certain constrains could be imposed on such parameters.  Then,  we  approach  the behaviors of  the light   rays  near     such   cosmological   black holes by  computing the  deflection angle in terms  of  EMD gravity  theory parameters.  Precisely, we elaborate a graphical discussion  in terms of the involved moduli space.  We show that the behavior of the deflection angle under the influence of the cosmological constant differs from that in  the Reissner-Nordström-AdS black holes due to the new coupling between the charge and the cosmological constant. This effect is mitigated by incorporating the contribution of the rotation parameter. Consequently, for rotating cosmological EMD black holes, the contribution of the cosmological constant to the deflection angle becomes similar to that in ordinary AdS black holes. \\

The paper is organized as follows. In Section 2, we provide a concise review of the AdS black holes in EMD gravity  theory.  It  presents the rotating black hole metric derived from  the Newman-Janis formalism without complexification. In Section 3, we  examine the shadow characteristics of both  non-rotating and rotating black holes, and  make contact with EHT empirical findings based on observational data.  Section 4  concerns the study  of  the deflection angle near  such EMD AdS  black holes. The last section is devoted to conclusions. 
\section{AdS  black holes in EMD  theory }
In this section, we provide a concise discussion on AdS black holes obtained from EMD gravity  theory. The latter has been  largely investigated in connection with low-energy limits of certain non-trivial theories including superstring models \cite{Hirschmann, Yu}. A close inspection shows that this theory can be elaborated via the coupling between supergravity and Maxwell abelian theory. Indeed, 
the dynamics  of Einstein-Maxwell-scalar theory is described by the following action 
\begin{equation}
S= \int d^4 x \sqrt{-g}\left(R - V(\varphi)-(\nabla \varphi)^2-K(\varphi) F^2\right) .
\end{equation}
In the context of this discussion, $R$ represents the Ricci scalar curvature, $F^2$ denotes  the quantity $F_{\mu \nu} F^{\mu \nu}$ arising from the Maxwell field strength.  $\varphi$ is the dilaton scalar field and the term $V(\varphi)$  indicates  the corresponding  scalar potential.   $K(\varphi)$ stands for the coupling function between the scalar field and the Maxwell field. This action can recover certain known models. Taking $K(\varphi)=1$ and considering $V(\varphi)=2 \lambda$, the theory reduces to the Einstein-Maxwell theory with a cosmological constant $\lambda$. This could supply the Reissner-Nordstrom-de Sitter solution. Setting $K(\varphi)=e^{2 \varphi}$ and $V(\varphi)=0$, however,  one gets  the dilaton black hole solution.  Additionally,  considering  $K(\varphi)=e^{2 \varphi}$ and $
 V(\varphi)=\frac{\lambda}{3}\left(e^{2 \varphi}+4+e^{-2 \varphi}\right)$, 
it yields the dilaton black hole in a de Sitter universe \cite{Hirschmann, Yu}. Besides these solutions, one  could  also recover   other significant solutions by considering different forms associated with $K(\varphi)$ and $V(\varphi)$ scalar quantities.  Conventionally, the gravity theory and the corresponding black hole solutions are determined once the expressions of $K(\varphi)$ and $V(\varphi)$ are furnished. Roughly speaking, the  line element metric $ds^2$ is recalled to be 
\begin{equation}
ds^2=-f(r) d t^2+f(r)^{-1} d r^2+h(r)\left(d \theta^2+\sin ^2 \theta d \varphi^2\right),
\end{equation}
where $f(r)$ and $h(r)$ are radial functions. Explicit forms of such functions can be derived from the equations of motion by fixing the involved scalar functions. Solving such equations for EMD gravity backgrounds, we find 
\begin{equation}
\begin{aligned}
 f(r)&=1-\frac{2 M}{r}-\frac{1}{3} \lambda h(r) , \\
h(r)&=r\left(r-\frac{Q^2}{M}\right)
\end{aligned}
  \end{equation}
where   $M$ and $Q$ represent the mass and  the electric charge of the black hole, respectively.  $\lambda$ is a dimensionless parameter within the EMD gravity  theory, often interpreted as a cosmological constant. It is important to note that when $Q = 0$, the solutions reduce to the Schwarzschild-de Sitter black hole. In this paper, we will focus on negative values of $\lambda$, which correspond to EMD AdS black holes. \\
To obtain the rotating solution of such  black holes, we employ the Newman-Janis method without complexification, applied to a general static and spherically symmetric metric \cite{Azreg, Ainou, Bezdekova}. Indeed,  the line element of the  rotating EMD AdS black holes  can take the following form
\begin{equation}
ds^2=-\left( 1-\frac{\sigma}{H} \right) dt^2-\frac{2a^2 \sigma}{H}\sin^{2} \theta dtd\phi+ \left( h(r) +a^2+\frac{a^2 \sigma \sin^{2} \theta }{H}\right)\sin^{2} \theta d\phi^2 +\frac{H}{\Delta}dr^2+Hd\theta^2.
\label{mr}
\end{equation}
In this solution, one has used
\begin{eqnarray}
H &=&  H(r)=h(r)+a^2 \cos^2\theta\\
\Delta&=&  \Delta(r)= f(r)h(r)+a^2\\
\sigma&=& \sigma(r) = h(r)\left( 1-f(r)\right)  
\end{eqnarray}
where $a$ is the rotating spin  parameter. It is worth noting that this rotating metric was derived under conditions that ensure it is a physically acceptable solution of the field equations, where the energy-momentum source can be interpreted as an imperfect fluid rotating about the 
$z$-axis.

 Having elaborated  the non-rotating and   the rotating black hole solutions, we move now to investigate their  physical properties. A special emphasis  will be  put on optical behaviors. Other aspects could be approached using certain  appropriate formalisms.
\section{Shadow behaviors and empirical  constraints}
In this section, we investigate the  optical shadow behaviors of EMD AdS  black holes  and try to make contact with empirical findings via  the EHT international collaboration.  
\subsection{Shadows of   cosmological EMD black holes}
In this section, we investigate the optical aspects of the dilaton black holes in the (anti-) de Sitter universe by examining their shadow properties. As we are focusing on four-dimensional black holes, we will provide graphical representations of their shadows using one-dimensional real closed curves.  To do so,  we consider  the Hamilton-Jacobi equation
\begin{equation}
\frac{\partial \mathcal{S}}{\partial \sigma}=-\frac{1}{2} g^{\mu \nu} \frac{\partial \mathcal{S}}{\partial x^\mu} \frac{\partial \mathcal{S}}{\partial x^\nu} ,
\end{equation}
where $\mathcal{S}$ is the Jacobi action which can be expressed as
\begin{equation}
\mathcal{S} =  - Et + L\phi + S_r(r) + S_\theta(\theta).
\end{equation}
In this way,   \(E\) is the energy, \(L\) is the angular momentum, \(S_r(r)\) is a function of the radial coordinate \(r\), and \(S_\theta(\theta)\) is a function of the polar angle \(\theta\).

Considering first the non-rotating	EMD black holes in the  AdS space,  the conserved total energy and the conserved angular
momentum of the massless particle are given by
\begin{equation}
\begin{gathered}
E=-p_{t}=-f(r) \dot{t} \\
L=p_{\phi}=h(r) \sin ^2 \theta \dot{\phi} .
\end{gathered}
\label{el}
\end{equation}
The relevant null geodesic equations of $r$ and $\theta$ coordinates can be derived using a separation method similar to the one employed in the Carter mechanism \cite{A81}. Indeed, we  can get
\begin{equation}
\begin{aligned}
&h(r)f(r) \left( \frac{d S_r(r)}{d r} \right)^2 + L^2 - \frac{f(r)}{f(r)}E^2 = -\mathcal{C}, \\
&\left( \frac{d S_\theta(\theta)}{d \theta} \right)^2 + L^2 \cot^2(\theta) = \mathcal{C},
\end{aligned}
\label{se}
\end{equation}
where $\mathcal{C}$ is a separable constant.  Using Eq.(\ref{el}) and Eq.(\ref{se}), the four equations of motion are found to be 
\begin{align}
\dot{t} &= -\frac{E}{f(r)} \\
\dot{r} &= \pm\sqrt{\mathcal{R}(r)} \\
h(r)\dot{\theta} &= \pm\sqrt{\mathcal{C}-L^2 \cot^2 \theta} \\
\dot{\phi} &= \frac{L}{h(r)\sin^2\theta}.
\end{align}
In this way,   the radial quantity $\mathcal{R}(r)$ is expressed as follows
\begin{equation}
\mathcal{R}(r)=E^2\left( 1-\frac{f(r)}{h(r)}(\zeta^2+\eta)\right) 
\end{equation}
 where $ \eta$  and  $\zeta$   are the dimensionless   impact parameters  given by  
\begin{equation}
\eta=\frac{\mathcal{C}}{E^2}, \quad \zeta=\frac{L}{E}.
\end{equation}
The shapes of the  black hole shadows can be illustrated through the study of  the unstable circular orbits. Specifically, this can be determined by taking the following conditions
\begin{equation}
\mathcal{R}(r) \bigg|_{r=r_o} = 0, \quad \left. \mathcal{R}_{\text{eff}}'(r) \right|_{r=r_o} = 0
\end{equation}
where $r_o$ denotes the circular orbit radius of the photon. Solving the above equation  system, one can obtain
\begin{eqnarray}
r_o&=&\frac{6 M^2+Q^2+\sqrt{36 M^4-20 M^2 Q^2+Q^4}}{4 M} \\
\zeta^2+\eta &=&  {\frac{ h(r_{o})}{f(r_{o})}} .
\end{eqnarray}
The apparent shape of the shadow of the black hole,  in such a space-time, can be obtained using the celestial
coordinates $\alpha$ and $\beta$. According to \cite{A81}, these coordinates can be expressed as follows
\begin{equation}
\begin{aligned}
\alpha & =\lim _{r_{ob } \rightarrow+\infty}\left(-r_{ob }^2 \sin \theta_{ob } \frac{d \phi}{d r}\right) \\
\beta & =\lim _{r_{ob } \rightarrow+\infty}\left(r_{ob }^2 \frac{d \theta}{d r}\right),
\end{aligned}
\end{equation}
where $r_{ob }$ is the distance of the observer from the black hole and $\theta_{ob }$ indicates the angle of
inclination between the line of the observer and the axis of rotation of the black hole. The
coordinate $\alpha$ represents the apparent orthogonal distance of the observed image from the axis of symmetry, while the coordinate $\beta$ denotes the apparent perpendicular distance of
the image from its projection onto the equatorial plane. Considering the null geodesics, one gets the shadow  equation which can be expressed as follows
\begin{equation}
     \alpha^2+\beta^2=\zeta^2 + \eta .
\end{equation}

\begin{figure}[t] 
		\begin{center}
		\centering
			\begin{tabbing}
			\centering
			\hspace{8.cm}\=\kill
			\includegraphics[scale=.7]{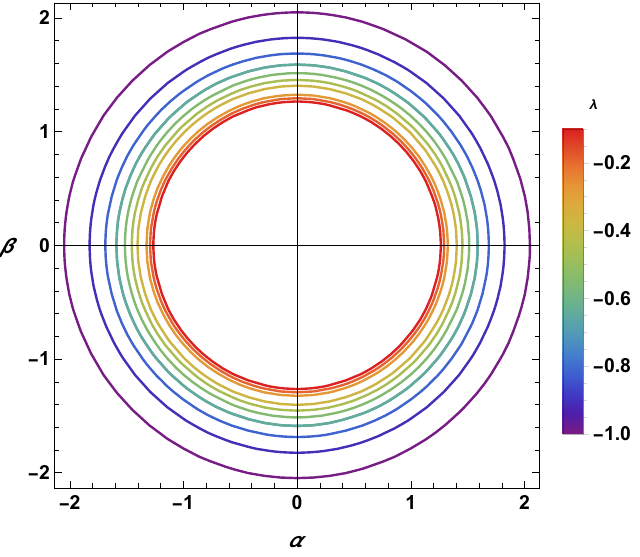} \>
			\includegraphics[scale=.7]{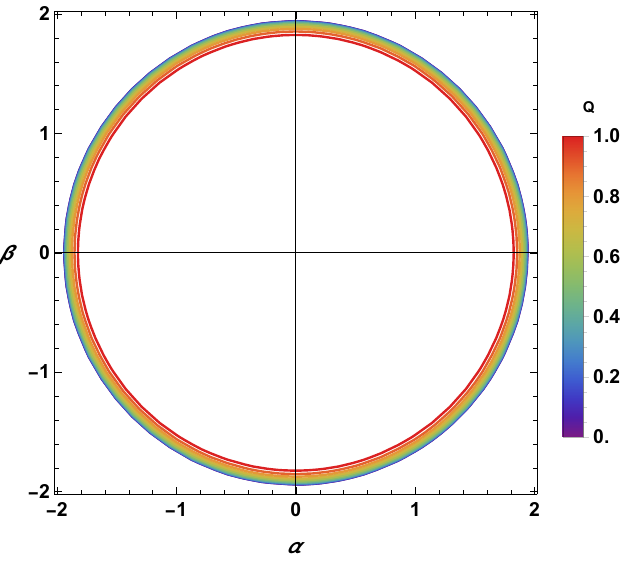} \\
		   \end{tabbing}
\caption{{\it \footnotesize Shadows of non-rotating EMD black holes. Left panel: $Q=M=1$, right panel: $M=1$ and $\lambda=-0.1$.}}
\label{SF1}
		   \end{center}
\end{figure}

In the Fig.(\ref{SF1}), we illustrate the shadow behaviors of the non-rotating  EMD  AdS black holes.  To examine the effects of  the  electric   charge and the cosmological constant in such  black hole models, we plot the shadow profile in the left panel, keeping the charge constant while varying the cosmological constant. In the right panel, we vary the charge while holding the cosmological constant constant.  As expected,  it  has been observed that the shadows are perfect circles, where their sizes are controlled by the parameters of such  black holes. Even though the form of the metric differs from that of the AdS Reissner-Nordström black holes, both the charge and the cosmological constant still act as decreasing parameters for the shadow radius. We conclude that these  parameters could be considered as size parameters.

\begin{figure}[!ht]
		\begin{center}
		\centering
			\begin{tabbing}
			\centering
			\hspace{2.cm}\=\kill
			\includegraphics[scale=0.58]{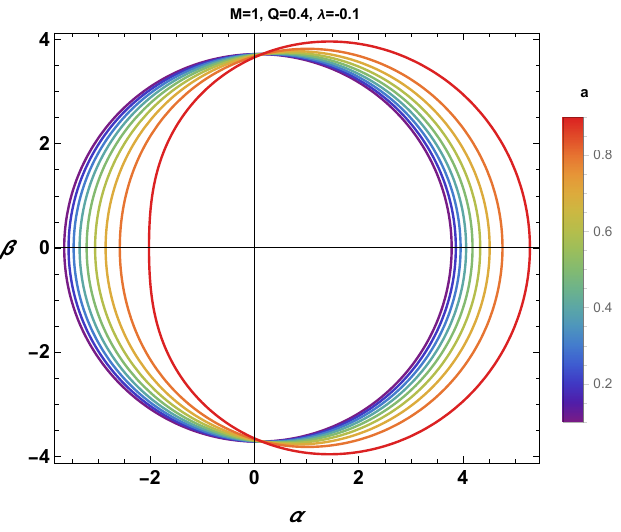} 
	\hspace{0.1cm}		\includegraphics[scale=0.5]{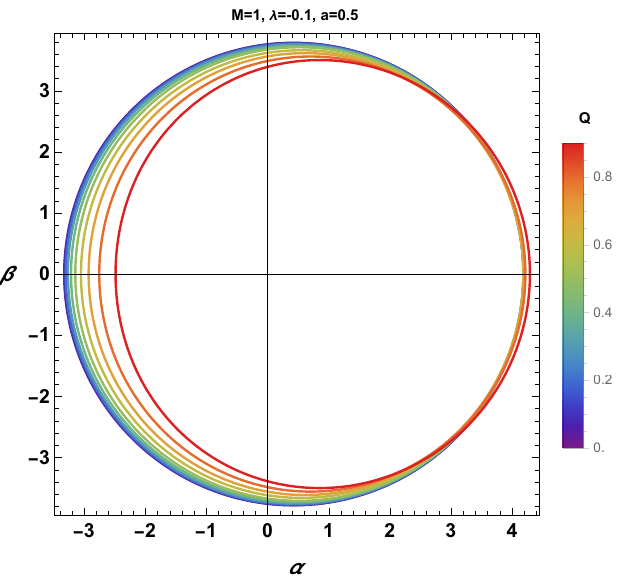}\hspace{0.1cm}	\includegraphics[scale=0.5]{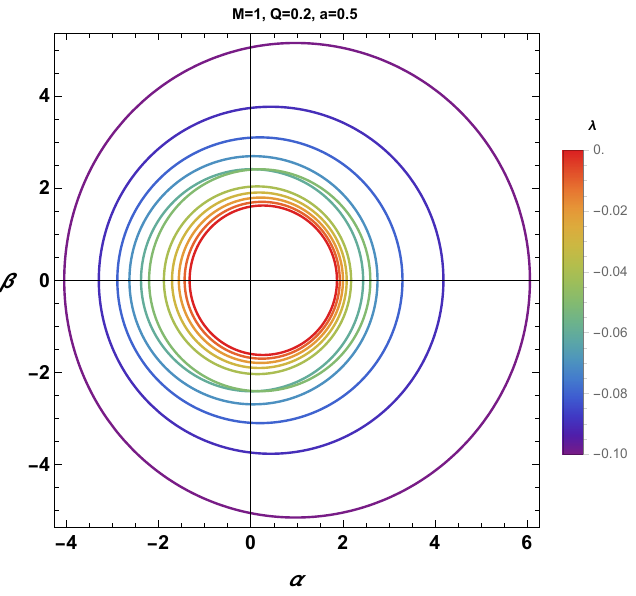}\\ 
	
		   \end{tabbing}
\caption{ \it \footnotesize  Shadow behaviors of AdS rotating EMD black hole in terms of the charge, the cosmological constant and the rotating parameter.}
\label{rsa}
\end{center}
\end{figure}

We move  now to approach  the    rotating EMD  AdS black hole solutions. Considering the metric (\ref{mr}) and applying the Hamilton Jacobi separation method, we get the following four equations of motion
\begin{align}
H  \dot{t}& = \frac{h+a^2}{\Delta}\left[ E\left( h+a^2\right) -a L\right] +a\left[  L-aE\sin^2\theta \right]  \\
(H \dot{r})^2 &=\mathcal{R}(r) \\
( H\dot{ \theta})^2 & =\Theta(\theta)\\
H  \dot{\phi} &= \left[ L \csc ^2 \theta-aE\right] +\frac{a}{\Delta} \left[ E\left(h+a^2\right) -aL\right],
\end{align}
where $ \mathcal{R}(r)$  and $\Theta(\theta) $ functions  are expressed as follows
 \begin{align}
\mathcal{R}(r) &= \left[ E \left( h + a^2 \right) - a L \right]^2 - \Delta \left[ \mathcal{C} + \left( L - a E \right)^2 \right], \\
\Theta(\theta) &= \mathcal{C} - \left( L \csc \theta - a E \sin \theta \right)^2 + \left( L - a E \right)^2.
\end{align}
Solving the  unstable circular orbit equations, the two  impact parameters   are obtained as follows 
\begin{align}
& \eta =\frac{4 \Delta (r) h'(r) \left(a^2 h'(r)+h(r) \Delta '(r)\right)-4 \Delta (r)^2 h'(r)^2-h(r)^2 \Delta '(r)^2}{a^2 \Delta '(r)^2}\vert_{r=r_0},\\
& \xi=\frac{a^2 \Delta '(r)-2 \Delta (r) h'(r)+h(r) \Delta '(r)}{a \Delta '(r)}\vert_{r=r_0}.
\end{align}
For the rotating EMD  AdS black holes, the apparent shapes of the shadows   are depicted in Fig.(\ref{rsa}) by considering the celestial
coordinates $\alpha$ and $\beta$.

This figure provides  the shadow configurations for certain black hole parameters.  It has been observed various shapes and sizes depending on the rotating parameter.  For small values of such a parameter, the  shadows are    perfect cercles  confirming the  above result associated with the  non-rotating solutions.  Augmenting  the values of  $a$, however,  we get deformed  geometries.  Precisely,  we observe that  the perfect cercles have  been distorted and deviated from the center  by varying   the remaining black hole parameters for certain  acceptable ranges. As expected, the D-shape has been obtained by considering  large values of the spin parameter.  Other behaviors have been observed. Increasing  the cosmological parameter and the charge,    the size of the obtained shadows decreases.
 \subsection{Constraints from empirical results}
 To approach an alignment between the rational  predictions and the empirical  data, the following subsection will present an analysis of the shadow cast by rotating cosmological EMD black holes, incorporating relevant data from the EHT collaboration. Specifically, we exploit  the EHT observational results for M$87^*$ and Sgr$ A^*$ to establish constraints on the parameters of these black holes \cite{
EventHorizon,Gogoi,Chakhchi}. { Using the fractional deviation from the Schwarzschild black hole shadow diameter, defined as
\begin{equation}
\delta=\frac{R_s}{r_{sh}},
\end{equation}
it has been demonstrated that these constraints can be imposed on black hole parameters through the dimensionless quantity \( R_s/M \)} where $R_s$ is shadow radii and $M$ is  the black hole mass. Indeed, the 1-\(\sigma\) and  the 2-\(\sigma\) measurements derived from observational data are provided in Table~\ref{t1}.

\begin{table}[h!]
\centering
\begin{tabular}{|c|c|c|c|}
\hline
\textbf{Black Hole} & \textbf{Deviation (\(\delta\))} & \textbf{1-\(\sigma\) Bounds} & \textbf{2-\(\sigma\) Bounds} \\ 
\hline
M87$^*$ (EHT) & $-0.01^{+0.17}_{-0.17}$ & $4.26 \leq \frac{R_s}{M} \leq 6.03$ & $3.38 \leq \frac{R_s}{M} \leq 6.91$ \\ 
\hline
Sgr A$^*$ (EHT$_{\text{VLTI}}$) & $-0.08^{+0.09}_{-0.09}$ & $4.31 \leq \frac{R_s}{M} \leq 5.25$ & $3.85 \leq \frac{R_s}{M} \leq 5.72$ \\ 
\hline
Sgr A$^*$ (EHT$_{\text{Keck}}$) & $-0.04^{+0.09}_{-0.10}$ & $4.47 \leq \frac{R_s}{M} \leq 5.46$ & $3.95 \leq \frac{R_s}{M} \leq 5.92$ \\ 
\hline
\end{tabular}
\caption{Estimates and bounds for M87$^*$ and Sgr A$^*$ black holes.}
\label{t1}
\end{table}

 \begin{figure}[!ht]
		\begin{center}
		\centering
			\begin{tabbing}
			\centering
			\hspace{6.cm}\=\kill
			\includegraphics[scale=0.73]{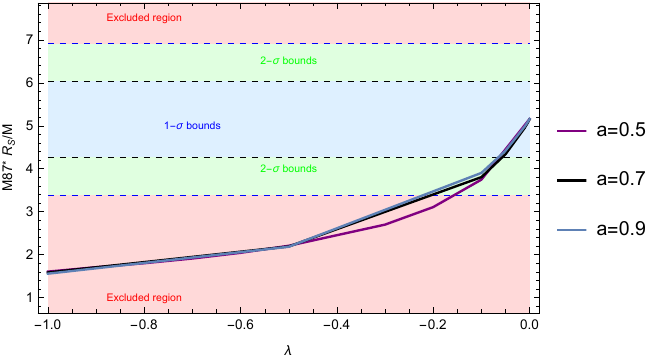} 
	\hspace{0.1cm}		\includegraphics[scale=0.73]{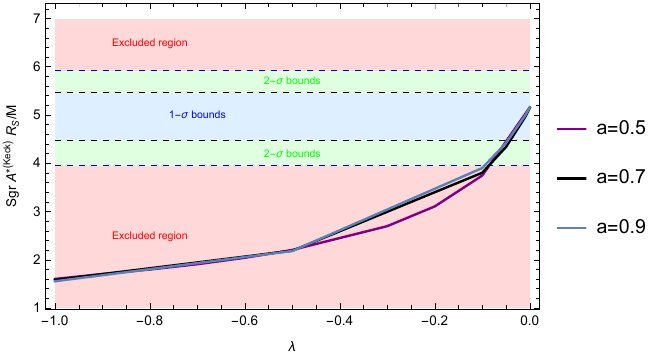}\hspace{0.1cm}	\\ 
	
		   \end{tabbing}
		   \includegraphics[scale=0.73]{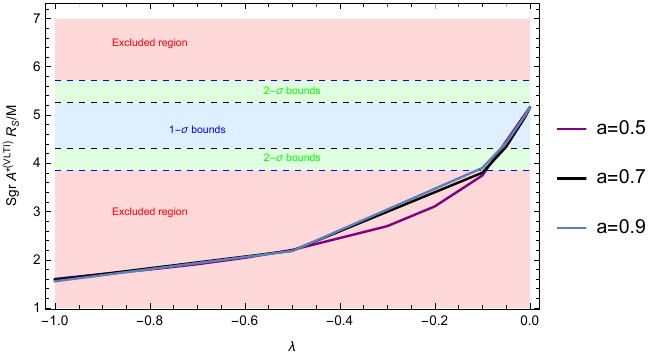}
\caption{ \it \footnotesize Regions of cosmological rotating EMD black holes that are $(1-\sigma)$ and $(2-\sigma)$ consistent or inconsistent with the EHT data, as a function of the cosmological constant. The plot shows the ratio $R_s/M$ for three values of the rotation parameter, with the charge $Q=0.2$ and mass $M=1$ .}
\label{Cs1}
\end{center}
\end{figure}

In Fig.(\ref{Cs1}), we illustrate the regions of cosmological rotating EMD AdS  black holes that are $(1-\sigma)$ and $(2-\sigma)$ consistent or inconsistent with the EHT data. These regions are shown by varying the cosmological constant for three different values of the rotation parameter. {  Based on the figure, we can deduce that for all three rotation parameter values, achieving consistency with the EHT data requires a very small negative value of the cosmological constant}. As the cosmological constant approaches this small negative value, the shadow radii align  closely with the observational constraints from EHT for M87* and Sgr A*.  { The variation of the cosmological constant in this negative regime plays a primordial role in fine-tuning the theoretical predictions to remain within the $(1-\sigma)$ and $(2-\sigma)$ confidence levels set by EHT observations}.

\begin{figure}[!ht]
		\begin{center}
		\centering
			\begin{tabbing}
			\centering
			\hspace{6.cm}\=\kill
			\includegraphics[scale=0.73]{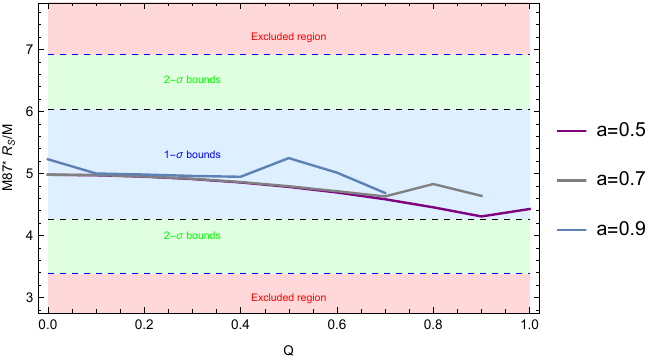} 
	\hspace{0.1cm}		\includegraphics[scale=0.73]{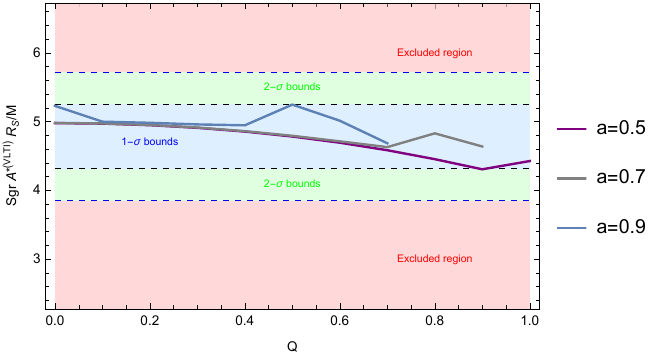}\hspace{0.1cm}\\ 
		   \end{tabbing}
		   	\includegraphics[scale=0.73]{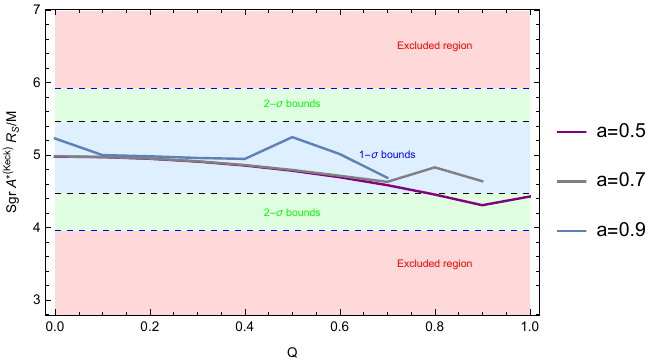}
\caption{ \it \footnotesize  Regions of cosmological rotating EMD black holes that are $(1-\sigma)$ and $(2-\sigma)$ consistent or inconsistent with the EHT  empirical data, as a function of the charge. The plot shows the ratio $R_s/M$ for three values of the rotation parameter, with the cosmological constant $\Lambda=0.01$ and mass $M=1$ .}
\label{Cs2}
\end{center}
\end{figure}

Indeed, by taking a small value of the cosmological constant, for instance, $\lambda = -0.01$, we now consider the variation of the shadow radii in terms of the charge. It is clear from Fig.(\ref{Cs2}) that a more consistent situation is presented. For different values of the electric  charge, the curves remain within the $(1-\sigma)$ bound, and for Sgr A* (EHT$_{\text{Keck}}$). The curves could  be shifted to the $(2-\sigma)$ bound for small values of the rotation parameter and large values of the charge.

 \section{Deflection angle computations}
 
 In this section, we approach  the behavior of the deflection of light rays by EMD AdS black holes. To calculate the deflection angle, we could adopt the approach based on the Gauss-Bonnet
theorem. The latter is one of the important method to compute the weak deflection angle   using the optical
geometry proposed by Gibbons and Werner \cite{A117,A118,A119}. Taking into account
that both the observer  ($R$) and the source ($S$) are located at finite distances within the equatorial plane,
the deflection angle can be formulated as follows
\begin{equation}
\Theta=\Psi_{R}-\Psi_{S}+\phi_{SR}
\label{a3}
\end{equation}
where  $\Psi_{R}$ and $\Psi_{S}$ represent the angles between the light rays and the radial direction at the positions of the observer and the source, respectively.

\begin{figure}[t] 
		\begin{center}
		\centering
			\begin{tabbing}
			\centering
			\hspace{9.cm}\=\kill
			\includegraphics[scale=.8]{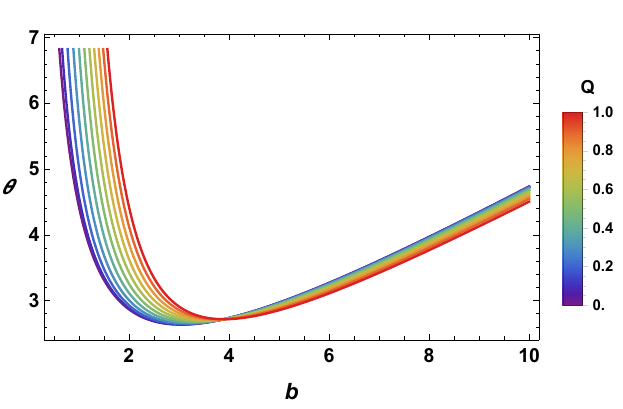} \>
			\includegraphics[scale=.8]{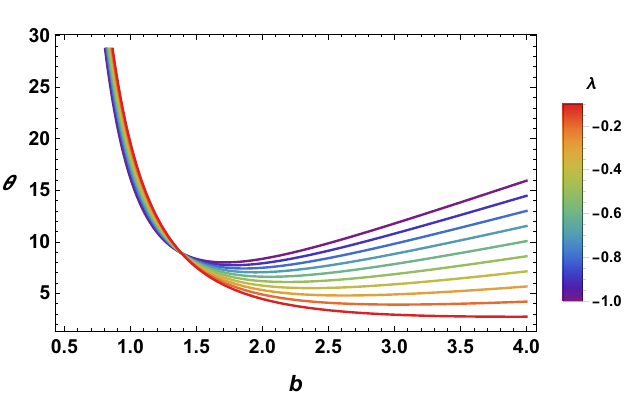} \\
		   \end{tabbing}
\caption{{\it \footnotesize Deflection angle of non rotating EMD  AdS black holes.  Right panel: $Q=M=1$,  left panel: $M=1$ and $\lambda=-0.1$.}}
\label{daF1}
		   \end{center}
\end{figure}

The angle $\phi_{SR}$ denotes the longitudinal separation between these positions, as described in \cite{A117}. To calculate these optical quantities, we can employ the algorithm developed in \cite{A118,A119}.  The separation angle can be computed using the following equation
\begin{equation}
\phi_{RS} = \int^R_S d\phi= \int^{u_0}_{u_S}\frac{1}{\sqrt{F(u)}}du +\int^{u_0}_{u_R}\frac{1}{\sqrt{F(u)}}du
\end{equation} 
where $u_{S}$ and $u_{R}$ are the inverse of the distances from the black hole to the source and the observer, respectively. The parameter $u_{0}$  indicates the inverse of the closest approach distance $r_{0}$ and $b$ is associated  with  the  impact parameter $\frac{L}{E}$. The function $F(u)$ is expressed as follows
\begin{eqnarray}
F(u)=\left(\frac{1}{u^2} \frac{du}{d\phi}\right)^2. 
\end{eqnarray}
Considering the metric of a non-rotating EMD black hole in an AdS background and taking the  follows order $\mathcal{O}(M^1,Q^2,\lambda^1)$, 
the computations lead to
{\small 
 \begin{multline}
\phi_{RS}=\left(\pi-\arcsin \left(b u_R\right)-\arcsin \left(b u_S\right) \right)+\left(\frac{2-b^2 u_R^2}{\sqrt{1-b^2 u_R^2}}+\frac{2-b^2 u_S^2}{\sqrt{1-b^2 u_S^2}}\right)\frac{G M}{b}\\- \left(\frac{u_R}{\sqrt{1-b^2 u_R^2}}+\frac{u_S}{\sqrt{1-b^2 u_S^2}}\right)\frac{b^3 \lambda}{6} +\left(\frac{3 b^2 u_R^2-2}{\left(1-b^2 u_R^2\right)^{3/2}}+\frac{3 b^2 u_S^2-2}{\left(1-b^2 u_S^2\right)^{3/2}}\right) \frac{b \lambda  M}{6}\\ +  \left(\frac{u_R^2}{\sqrt{1-b^2 u_R^2}}+\frac{u_S^2}{\sqrt{1-b^2 u_S^2}}\right)\frac{b Q^2 }{2 M}-\left(\frac{u_R \left(b^4 u_R^4+4 b^2 u_R^2-3\right)}{\left(1-b^2 u_R^2\right)^{3/2}}+\frac{u_S \left(b^4 u_S^4+4 b^2 u_S^2-3\right)}{\left(1-b^2 u_S^2\right)^{3/2}}\right) \frac{Q^2 }{4 b}\\+ \left(\frac{b^8 u_R^8-10 b^6 u_R^6+60 b^4 u_R^4-80 b^2 u_R^2+32}{\left(1-b^2 u_R^2\right)^{5/2}}+\frac{b^8 u_S^8-10 b^6 u_S^6+60 b^4 u_S^4-80 b^2 u_S^2+32}{ \left(1-b^2 u_S^2\right)^{5/2}}\right)\frac{ M Q^2}{4 b^3}\\-\left(\frac{u_R^2}{\left(1-b^2 u_R^2\right)^{3/2}}+\frac{u_S^2}{\left(1-b^2 u_S^2\right)^{3/2}}\right) \frac{ b^3 \lambda  Q^2}{12 M}
 + \left(\frac{u_R \left(7 b^4 u_R^4-7 b^2 u_R^2+3\right)}{\left(1-b^2 u_R^2\right)^{5/2}}+\frac{u_S \left(7 b^4 u_S^4-7 b^2 u_S^2+3\right)}{\left(1-b^2 u_S^2\right)^{5/2}}\right) \frac{b \lambda  Q^2}{12}
 \\+ \left(\frac{5 b^8 u_R^8-70 b^6 u_R^6+140 b^4 u_R^4-112 b^2 u_R^2+32}{\left(1-b^2 u_R^2\right)^{7/2}}+ \frac{5 b^8 u_S^8-70 b^6 u_S^6+140 b^4 u_S^4-112 b^2 u_S^2+32}{\left(1-b^2 u_S^2\right)^{7/2}}\right) \frac{ \lambda  M Q^2}{8 b}.
\end{multline}}
  According to \cite{A117,A118,A119},     we  could get  the $\psi$ terms. Indeed,  we find 
 \begin{eqnarray}
\Psi_{R}&-&\Psi_{S}=\left(\arcsin \left(b u_R\right)+\arcsin \left(b u_S\right)-\pi \right)-\left(\frac{u_R^2}{\sqrt{1-b^2 u_R^2}}+\frac{u_S^2}{\sqrt{1-b^2 u_S^2}}\right)b  M \notag 
\\&-&   \left(\frac{1}{u_R \sqrt{1-b^2 u_R^2}}+\frac{1}{u_S \sqrt{1-b^2 u_S^2}}\right)\frac{b \lambda}{6} - \left(\frac{1-2 b^2 u_{R}^2}{\left(1-b^2 u_{R}^2\right)^{3/2}}+\frac{1-2 b^2 u_{S}^2}{\left(1-b^2 u_{S}^2\right)^{3/2}}\right) \frac{b \lambda  M}{6} \notag
\\&+& \left(\frac{u_R ^2}{\sqrt{1-b^2 u_R ^2}}+\frac{u_S ^2}{\sqrt{1-b^2 u_S^2}}\right) \frac{b Q^2 }{2 M}- \left(\frac{u_R^3}{\left(1-b^2 u_R^2\right)^{3/2}}+\frac{u_S^3}{\left(1-b^2 u_S^2\right)^{3/2}}\right) \frac{b Q^2}{2} \notag\\&-&  \left(\frac{u_R^3 \left(b^5 u_R^5-5 b^3 u_R^3+b u_R\right)}{\left(1-b^2 u_R^2\right)^{5/2}}+\frac{u_S^3 \left(b^5 u_S^5-5 b^3 u_S^3+b u_S\right)}{\left(1-b^2 u_S^2\right)^{5/2}}\right) \frac{M Q^2}{4}\notag \\
&-&\left(\frac{2 b^2 u_R^2-1}{\left(1-b^2 u_R^2\right)^{3/2}}+\frac{2 b^2 u_S^2-1}{\left(1-b^2 u_S^2\right)^{3/2}}\right)\frac{ b \lambda  Q^2}{12 M} \\
&+& \left(\frac{4 b^5 u_R^5-2 b^3 u_R^3+b u_R}{\left(1-b^2 u_R^2\right)^{5/2}}+\frac{4 b^5 u_S^5-2 b^3 u_S^3+b u_S}{\left(1-b^2 u_S^2\right)^{5/2}}\right) \frac{\lambda  Q^2}{12}\notag\\
&-& \left(\frac{u_R \left(10 b^7 u_R^7+b^5 u_R^5+7 b^3 u_R^3-3 b u_R\right)}{\left(1-b^2 u_R^2\right)^{7/2}}+\frac{u_S \left(10 b^7 u_S^7+b^5 u_S^5+7 b^3 u_S^3-3 b u_S\right)}{\left(1-b^2 u_S^2\right)^{7/2}} \right) \frac{\lambda  M Q^2}{24}.\notag
\end{eqnarray}
Combining the above equations, we  can get the expression of the light deflection angle  by  the non-rotating EMD black holes in the AdS backgrounds. Precisely, it has been found to be 
{\small
\begin{multline}
\Theta =\left(\sqrt{1-b^2 u_R^2}+\sqrt{1-b^2 u_S^2}\right)\frac{2 M }{b}-  \left(\frac{b^2 u_R^2+1}{u_R \sqrt{1-b^2 u_R^2}}+\frac{b^2 u_S^2+1}{u_S \sqrt{1-b^2 u_S^2}}\right)\frac{b \lambda }{6}\\+ \left(\frac{5 b^2 u_R^2-3}{\left(1-b^2 u_R^2\right)^{3/2}}+\frac{5 b^2 u_S^2-3}{\left(1-b^2 u_S^2\right)^{3/2}}\right)\frac{b M \lambda }{6} +\left(\frac{u_R^2}{\sqrt{1-b^2 u_R^2}}+\frac{u_S^2}{\sqrt{1-b^2 u_S^2}}\right)\frac{b Q^2}{ M}\\- \left(\frac{u_R \left(b^4 u_R^4+6 b^2 u_R^2-3\right)}{\left(1-b^2 u_R^2\right)^{3/2}}+\frac{u_S \left(b^4 u_S^4+6 b^2 u_S^2-3\right)}{\left(1-b^2 u_S^2\right)^{3/2}}\right)\frac{Q^2 }{4 b}- \left(\frac{3 b^2 u_R^2-1}{\left(1-b^2 u_R^2\right)^{3/2}}+\frac{3 b^2 u_S^2-1}{\left(1-b^2 u_S^2\right)^{3/2}}\right) \frac{b \lambda  Q^2}{12 M}\\+\left(\frac{-5 b^6 u_R^6+59 b^4 u_R^4-80 b^2 u_R^2+32}{\left(1-b^2 u_R^2\right)^{5/2}}+\frac{-5 b^6 u_S^6+59 b^4 u_S^4-80 b^2 u_S^2+32}{\left(1-b^2 u_S^2\right)^{5/2}}\right) \frac{M Q^2}{4 b^3}\\+\left(\frac{5 b^8 u_R^8-211 b^6 u_R^6+413 b^4 u_R^4-333 b^2  u_R^2+96}{\left(1-b^2 u_R^2\right)^{7/2}}+\frac{5 b^8 u_S^8-211 b^6 u_S^6+413 b^4 u_S^4-333 b^2 u_S^2+96}{\left(1-b^2 u_S^2\right)^{7/2}}\right)\frac{ \lambda  M Q^2}{24 b}
\\+ \left(\frac{u_R \left(11 b^4 u_R^4-9 b^2 u_R^2+4\right)}{\left(1-b^2 u_R^2\right)^{5/2}}+\frac{u_S \left(11 b^4 u_S^4-9 b^2 u_S^2+4\right)}{\left(1-b^2 u_S^2\right)^{5/2}}\right) \frac{b \lambda  Q^2}{12}.
\end{multline}}
It has been noted that the above expression diverges as the limits \( bu_S \to 0 \) and \( bu_R \to 0 \) are approached. This divergence arises from the presence of the cosmological constant. Consequently, the finite deflection angle of light rays by AdS black holes  in EMD gravity theory  reads  approximately as 
\begin{eqnarray}
\theta \simeq \frac{4 M}{b}+\frac{16 M Q^2}{b^3} -\frac{b \lambda }{6}  \left(\frac{1}{{u_R}}+\frac{1}{{u_S}}\right)-b \lambda  M+\frac{8 \lambda  M Q^2}{b}+\frac{b \lambda  Q^2}{6 M}.
\end{eqnarray}
As expected, the deflection angle depends on the black hole  parameters, including $M$, $b$, $Q$, and $\lambda$. The term corresponding to the Schwarzschild deflection angle has been  recovered. Conversely, the contributions of the other parameters in the expression have been altered due to their coupling in the metric.    To  examine such optical quantity,   Fig.(\ref{daF1}) represents  the deflection angle  as a function of the impact parameter $b$ by considering   different values of   $Q$ and $\lambda$.  

Taking small values of $b$, we observe that the deflection angle initially decreases as a function of the impact parameter until a turning point, where the deflection becomes an increasing function of the impact parameter. As mentioned in  our previous works \cite{BelhajH}, this behavior is linked to the presence of the cosmological constant. Moreover, the charge acts as an increasing parameter of the deflection angle  before reaching the  turning point and then it becomes a decreasing parameter. The same behavior has been observed in the case of  the deflection angle of the  RN-AdS black hole.
Moreover, the charge initially acts as an increasing parameter for the deflection angle until reaching a turning point, after which it becomes a decreasing parameter. A similar behavior is observed in the deflection angle of the RN-AdS black hole. The coupling between the charge and the cosmological constant in the EMD AdS black hole leads to a change in the deflection angle  behavior under variations of the cosmological constant. For the RN-AdS black holes, the cosmological constant behaves  like a decreasing parameter. However, in this case, there is an initial range of the impact parameter where the deflection angle decreases with 
$\lambda$, and then it becomes an increasing parameter.

Now, we investigate the effect of the rotation parameter on the deflection angle of the  light rays by rotating  EMD AdS  black holes. By considering the metric (\ref{mr}) and applying the previously established formalism, we get  the separation angle  for the order   $\mathcal{O}(M^1,Q^2,\lambda^1,a^1)$ as follows

 \begin{eqnarray}
\phi_{RS_a}&=&\phi_{RS}+ \left(\frac{2 b^2 u_R^2-1}{u_R\sqrt{1-b^2 u_R^2}}+\frac{2 b^2 u_S^2-1}{u_S \sqrt{1-b^2 u_S^2}}\right)\frac{a \lambda}{3} +\left(\frac{1}{\sqrt{1-b^2 u_R^2}}+\frac{1}{\sqrt{1-b^2 u_S^2}}\right) \frac{2 a M}{b^2}\notag \\
&+&  \left(\frac{1-2 b^2 u_R^2}{\left(1-b^2 u_R^2\right)^{3/2}}+\frac{1-2 b^2 u_S^2}{\left(1-b^2 u_S^2\right)^{3/2}}\right) \frac{2 a \lambda  M}{3}+ \left(\frac{1}{\left(1-b^2 u_R^2\right)^{3/2}}+\frac{1}{\left(1-b^2 u_S^2\right)^{3/2}}\right)\frac{ a \lambda  Q}{6 M}\notag \\&+&  \left(\frac{u_R \left(2 b^2 u_R^2-1\right)}{\left(1-b^2 u_R^2\right)^{3/2}}+\frac{u_S \left(2 b^2 u_S^2-1\right)}{\left(1-b^2 u_S^2\right)^{3/2}}\right) \frac{a Q^2}{b^2}-\left(\frac{u_R^5}{\left(1-b^2 u_R^2\right)^{5/2}}+\frac{u_S^5}{\left(1-b^2 u_S^2\right)^{5/2}}\right)a b^4 \lambda  Q^2 \notag \\&-&\left(\frac{-5 b^6 u_R^6+44 b^4 u_R^4-60 b^2 u_R^2+24}{\left(1-b^2 u_R^2\right)^{5/2}}+\frac{-5 b^6 u_S^6+44 b^4 u_S^4-60 b^2 u_S^2+24}{\left(1-b^2 u_S^2\right)^{5/2}}\right)\frac{a M Q^2}{b^4}\\&+&\left(\frac{35 b^6 u_R^6-70 b^4 u_R^4+56 b^2 u_R^2-16}{\left(1-b^2 u_R^2\right)^{7/2}}+\frac{35 b^6 u_S^6-70 b^4 u_S^4+56 b^2 u_S^2-16}{\left(1-b^2 u_S^2\right)^{7/2}}\right) \frac{3 a \lambda  M Q^2}{4 b^2}.\notag 
\end{eqnarray}
The $\psi$ term takes the following form
\begin{eqnarray}
\Psi_{R_a}&-&\Psi_{S_a}=\Psi_{R}-\Psi_{S}+ \left(\frac{1}{u_R \sqrt{1-b^2 u_R^2}}+\frac{1}{u_S \sqrt{1-b^2 u_S^2}}\right)\frac{a \lambda}{3} \notag\\&+& \left(\frac{u_R^2}{\sqrt{1-b^2 u_R^2}}+\frac{u_S^2}{\sqrt{1-b^2 u_S^2}}\right)2 a M+ \left(\frac{1-2 b^2 u_R^2}{\left(1-b^2 u_R^2\right)^{3/2}}+\frac{1-2 b^2 u_S^2}{\left(1-b^2 u_S^2\right)^{3/2}}\right)\frac{2 a \lambda  M}{3}\notag\\&-& \left(\frac{1-2 b^2 u_R^2}{\left(1-b^2 u_R^2\right)^{3/2}}+\frac{1-2 b^2 u_S^2}{\left(1-b^2 u_S^2\right)^{3/2}}\right)\frac{a \lambda  Q^2}{6 M}+ \left(\frac{u_R^3}{\left(1-b^2 u_R^2\right)^{3/2}}+\frac{u_S^3}{\left(1-b^2 u_S^2\right)^{3/2}}\right)a Q^2\notag\\&+& \left(\frac{-4 b^4 u_R^5+2 b^2 u_R^3-u_R}{\left(1-b^2 u_R^2\right)^{5/2}}+\frac{-4 b^4 u_S^5+2 b^2 u_S^3-u_S}{\left(1-b^2 u_S^2\right)^{5/2}}\right) \frac{a \lambda  Q^2}{3}\\&+& \left(\frac{b^4 u_R^8-5 b^2 u_R^6+u_R^4}{\left(1-b^2 u_R^2\right)^{5/2}}+\frac{b^4 u_S^8-5 b^2 u_S^6+u_S^4}{\left(1-b^2 u_S^2\right)^{5/2}}\right)a M Q^2\notag\\&+& \left(\frac{u_R^2 \left(10 b^6 u_R^6+b^4 u_R^4+7 b^2 u_R^2-3\right)}{\left(1-b^2 u_R^2\right)^{7/2}}+\frac{u_S^2 \left(10 b^6 u_S^6+b^4 u_S^4+7 b^2 u_S^2-3\right)}{\left(1-b^2 u_S^2\right)^{7/2}}\right) \frac{a \lambda  M Q^2}{4}.\notag
\end{eqnarray}
Indeed, the expression of the deflection angle is found to be
{\small
\begin{multline}
\theta_{a}=\theta +\left(\frac{b^2 u_R^2+1}{\sqrt{1-b^2 u_R^2}}+\frac{b^2 u_S^2+1}{\sqrt{1-b^2 u_S^2}}\right)\frac{2 a M }{b^2}+ \left(\frac{u_R}{\sqrt{1-b^2 u_R^2}}+\frac{u_S}{\sqrt{1-b^2 u_S^2}}\right) \frac{2 a b^2 \lambda }{3}\\+\left(\frac{u_R^2}{\left(1-b^2 u_R^2\right)^{3/2}}+\frac{u_S^2}{\left(1-b^2 u_S^2\right)^{3/2}}\right)\frac{ a b^2 \lambda  Q^2}{3 M}+ \left(\frac{1-2 b^2 u_R^2}{\left(1-b^2 u_R^2\right)^{3/2}}+\frac{1-2 b^2 u_S^2}{\left(1-b^2 u_S^2\right)^{3/2}}\right)\frac{4 a\lambda  M}{3}\\-\left(\frac{u_R \left(1-3 b^2 u_R^2\right)}{\left(1-b^2 u_R^2\right)^{3/2}}+\frac{u_S \left(1-3 b^2 u_S^2\right)}{\left(1-b^2 u_S^2\right)^{3/2}}\right)\frac{a Q^2}{b^2}+ \left(\frac{-7 b^4 u_R^5+2 b^2 u_R^3-u_R}{\left(1-b^2 u_R^2\right)^{5/2}}+\frac{-7 b^4 u_S^5+2 b^2 u_S^3-u_S}{\left(1-b^2 u_S^2\right)^{5/2}}\right) \frac{a \lambda  Q^2}{3}\\+\left(\frac{b^8 u^8-43 b^4 u^4+60 b^2 u^2-24}{\left(1-b^2 u^2\right)^{5/2}}+\frac{b^8 u^8-43 b^4 u^4+60 b^2 u^2-24}{\left(1-b^2 u^2\right)^{5/2}}\right)\frac{a M Q^2}{b^4}+\frac{a \lambda  M Q^2}{4 b^2}\\ \left(\frac{10 b^8 u_R^8+106 b^6 u_R^6-203 b^4 u_R^4+165 b^2 u_R^2-48}{\left(1-b^2 u_R^2\right)^{7/2}}+\frac{10 b^8 u_S^8+106 b^6 u_S^6-203 b^4 u_S^4+165 b^2 u_S^2-48}{\left(1-b^2 u_S^2\right)^{7/2}}\right).
\end{multline} }
Considering the limits $bu_S \to 0$  and  $ bu_R \to 0$,   we obtain 
\begin{eqnarray}
\theta &\simeq & \frac{4 M}{b}+\frac{16 M Q^2}{b^3} -\frac{b \lambda }{6}  \left(\frac{1}{{u_R}}+\frac{1}{{u_S}}\right)-b \lambda  M+\frac{8 \lambda  M Q^2}{b}+\frac{b \lambda  Q^2}{6 M}+\frac{2 a M}{b^2}+\frac{4 a \lambda  M}{3}\notag \\&-&\frac{24 a M Q^2}{b^4}-\frac{12 a \lambda  M Q^2}{b^2}.
\end{eqnarray}

Setting $a$, we recover the expression obtained for the non-rotating solutions  being discussed  previously. By considering 
$\lambda=0$, we retrieve the expression for the deflection angle of rotating black holes. The differences in the coefficients of the charge and the cosmological constant are noted. To inspect the behavior of the deflection angle, Fig.(\ref{ben2}) illustrates its variation as a function of 
$
b$, with different values of the involved parameters.

\begin{figure}[t] 
		\begin{center}
		\centering
			\begin{tabbing}
			\centering
			\hspace{9.cm}\=\kill
			\includegraphics[scale=.8]{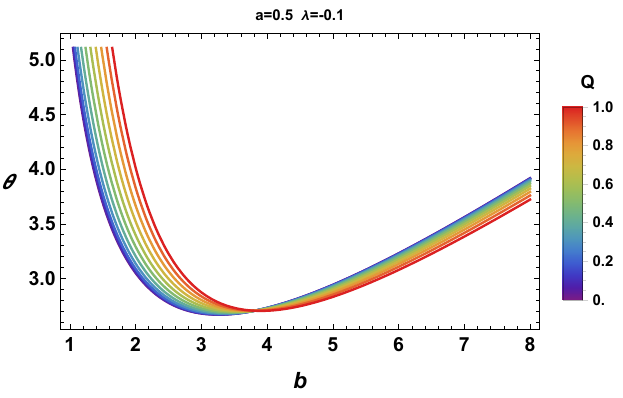} \>
			\includegraphics[scale=.8]{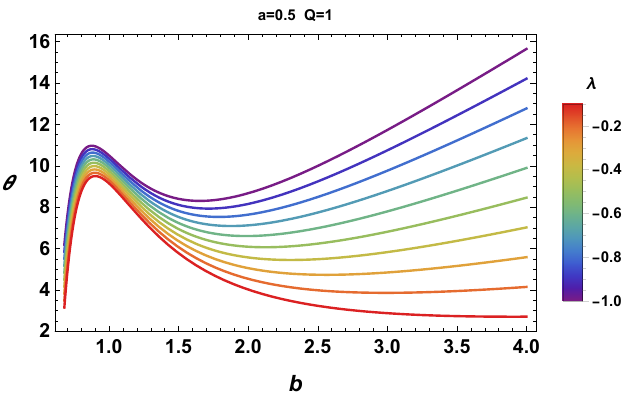} \\
		   \end{tabbing}
		   \includegraphics[scale=0.8]{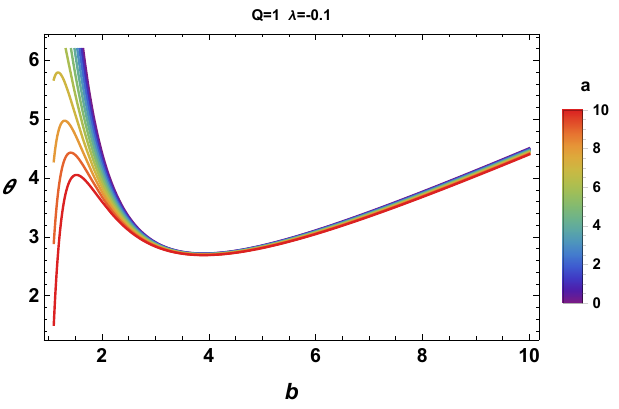}\\
\caption{\it \footnotesize Deflection angle of rotating EMD AdS black holes by taking $M=1$ and varying the remaining parameters.}
    \label{fig:ben2}
		   \end{center}
\end{figure}

The figure clearly shows that the effect of the charge remains consistent with that observed in the non-rotating black holes. The only difference is that the deflection angle values are lower when the rotation parameter is non-zero. Varying only the rotation parameter, it becomes evident that this parameter reduces the deflection angle. It is worth noting that this behavior perfectly matches   with the ones  observed in  the rotating AdS black holes. Interestingly, the inflection behavior observed in the first part is eliminated by the contribution of the rotation parameter, making the behavior of the deflection angle under changes in the cosmological constant similar to that of the AdS black holes.
\newpage
\section{Conclusion and open questions}
Motivated by  string theory models, we have studied certain behaviors of   cosmological black holes in EMD gravity theory.   Concretely, we have approached two relevant optical quantities: shadow and deflection angle.  Concerning    the shadows,  we  have first  investigated   the non-rotating   solutions.  As expected, we  have found    perfect circular shadows matching  with the previous results. Moreover,  we have  observed   that the  electric charge and the cosmological constant  behave like  decreasing parameters for the shadow radius. Using   the  Newman-Janis formalism and  the Hamilton-Jacobi algorithm, we  have approached  the rotating  black hole shadows  via    one dimensional real curves.  In particular,    we  have remarked that the   size and shape shadows  depend  on various parameter including    the rotating  one.  To  make contact with EHT observational  data,  we have revealed that  certain constrains should be  imposed on such parameters.   In the final part of this work, we have  examined the behaviors of  the light   rays  near    such   cosmological   black holes by  computing the  deflection angle in terms  of   EMD  theory parameters.  Concretely,  we have elaborated a graphical  discussion of such a deflection angle of the light rays for non-rotating  and rotating solutions.    We have shown that the behavior of the deflection angle under the influence of the cosmological constant differs from that of the  Reissner-Nordström-AdS black holes, due to the new coupling between the charge and the cosmological constant. This effect is mitigated by incorporating the contribution of the rotation parameter. Consequently, for rotating  EMD black holes, the cosmological constant contributes to the deflection angle in a manner similar to its contribution in  the Reissner-Nordström-AdS black holes.

This work comes up with certain open questions.  A natural question  is   either  to  enlarge the corresponding moduli space or to   consider  other scalar fields  originated from string theory via different scenarios \cite{hajarp}. Another possible road  is to make contact with geometric deformations of   stringy compactification spaces   including the Calabi-Yau spaces with lower values of the  Hodge numbers \cite{hajar4}. This could be addressed in future works.
\section*{Acknowledgements} Hajar Belmahi would like to thank  Adil Belhaj  for  collaboration  and encouragement.  She would like also  to thank him     for useful  comments and discussions    on the manuscript draft.

\end{document}